\documentclass[fleqn,10pt]{article}
\usepackage{amsmath}
\usepackage{latexsym}
\usepackage{rotate,epsfig}
\usepackage{lscape}
\usepackage{color}
\usepackage{amsthm}
\usepackage{amsfonts}
\usepackage{amsbsy}
\usepackage{latexsym}
\usepackage{mathabx}
\newtheorem{theorem}{\bf Theorem}[section]

\theoremstyle{definition}

\newtheorem{definition}[theorem]{\bf Definition}
\newtheorem{example}[theorem]{\bf Example }
\usepackage{color}
\newcommand{\rd}{\textcolor{black}}
\begin{document}
\title{\bf Redundancy schemes for engineering coherent systems via a signature-based approach}

\author{Mahdi Doostparast \footnote{Email address: doustparast@um.ac.ir\ \& doostparast@math.um.ac.ir}
\vspace{2mm}\\
{\it\small{ Department of Statistics, Faculty of Mathematical Sciences,}}
\vspace{-0.1cm}\\
{\it\small{Ferdowsi University of Mashhad, Mashhad,\ Iran }}}
\date{}
\maketitle

\begin{abstract}
This paper proposes a signature-based approach for solving redundancy allocation problems when component lifetimes are not only heterogeneous but also dependent. The two common schemes for allocations, that is active and standby redundancies, are considered. If the component lifetimes are independent, the proposed approach leads to simple manipulations. Various illustrative examples are also analysed. This method can be implemented for practical complex engineering systems.
\end{abstract}
%\tableofcontents
\section{Introduction}\label{intro}
Redundancy policies are usually used to increase the reliabilities of engineering systems, if exist. In common, there are two schemes to allocate redundant components to the system, called ``active'' and ``standby'' redundancy allocations. In the former, the spars are put in parallel to the components of the system while in the later, the spars start functioning immediately after component failures. Determination of an optimal redundancy allocation in engineering systems is great of interest.

There are many researches on the redundancy allocation problem (RAP) and deriving optimal  allocations. For example, Boland et al. \cite{Boland1992} considered some stochastic orders for a $k$-out-of-$n$ system, which works whenever at least $n-k + 1$ components among $n$ components work. They proved that the optimal active redundancy policy always allocates the spare to the weakest component in series systems if the component lifetimes are independent. For the standby redundancy and under the likelihood ratio ordering, Boland et al. \cite{Boland1992} provided also sufficient conditions in which for series systems, the spare should be allocated to the weakest component while in parallel systems, it allocates the spare component to the strongest one. For recent results on RAP, see Singh and Misra \cite{singmis1994}, Singh and Singh \cite{singsin1997}, Mi \cite{mi1999}, Valdes and Zequeira \cite{valdes2003}, Romera et al. \cite{romera2004}, Hu and Wang \cite{hu2008}, Jeddi and Doostparast \cite{jeddi2015} and references therein.

The concept of ``signature'' was introduced by Samaniego \cite{Samaniego1985} and it is a useful tool for analysing stochastic behaviours of systems from a theoretical view of point. Precisely, 
let $\mathbf{X}=(X_1,\cdots,X_n)$ stand for absolutely continuous component lifetimes in a coherent system of order $n$ and $T=\phi(\mathbf{X})$ is the system lifetime and $\phi(.)$ stands for the ``structure system function".
If the component lifetimes are independent and identically distributed (IID), then the system signature is the vector  $\mathbf{s}=(s_1,\cdots,s_n)$ where $s_i=P(T=X_{i:n})$ and $X_{i:n}$ denotes $i$-th order statistics among $X_1,\cdots,X_n$.
Samaniego \cite{Samaniego1985} shows that the system reliability function is given by $\bar{F}_T(t)=P(T>t)=\sum_{i=1}^n s_i \bar{F}_{i:n}(t)$ where $\bar{F}_{i:n}(t)=P(X_{i:n}>t)$ for $t>0$. This paper suggests a signature-based approach for RAP. Therefore, the rest of this paper is organized as follow: In Section 2, the proposed signature-based is proposed for systems with independent but heterogeneous component lifetimes.
The RAP for systems with dependent component lifetimes is also studied in Section 3. Finally, Section 4 concludes. Illustrative  examples are given throughout the paper.

\section{RAP with independent components}\label{Sec:Suggesting:Scheme}
In this section, we assume that the component lifetimes $X_1,\cdots,X_n$ are independent but heterogeneous with respective reliability functions $\bar{F}_1,\cdots,\bar{F}_n$, i.e.
\[\bar{F}_i(t)=P(X_i>t),\ \ \ \forall t>0, \ \ i=1,\cdots,n.\]
Let 
\begin{equation}\label{G-bar}
\bar{G}(t)=h^{-1}(H(\bar{F}_1(t),\cdots,\bar{F}_n(t))),\ \ \forall t>0,
\end{equation}
where $H(p_1,\cdots,p_n)$ is a multinomial expression, called ``the structure reliability function" and $h(p)=H(p,\cdots,p)$ is the diagonal section of $H(p_1,\cdots,p_n)$.
Navarro et al. \cite{Navarro2011} proved that the reliability function of the system lifetime, $\bar{F}_T(t)=P(T>t)$, can be expressed as
\begin{equation}\label{equivalent:G:bar}
\bar{F}_T(t)=H(\bar{F}_1(t),\cdots,\bar{F}_n(t))=\sum_{i=1}^n s_i \bar{G}_{i:n}(t),\ \ \ \forall\ t>0,
\end{equation}
where $\bar{G}_{i:n}(t)$ stands for the reliability function of the $i$-th order statistics on the basis of a random sample of size $n$ from the distribution function (DF) $G(t)=1-\bar{G}(t)$ for $t>0$; that is
\begin{eqnarray}
\bar{G}_{i:n}(t)&=&\sum_{j=0}^{i-1} {n \choose j} G(t)^j \bar{G}(t)^{n-j}\nonumber\\
&=&1-E(n,i,G(t)),\ \ \ t>0, \ \ i=1,\cdots,n. \label{G-bar-expression}
\end{eqnarray}
where $E(n,i,a)=\sum_{j=i}^n {n\choose j} a^j (1-a)^{n-j}$. 
In other words, for every given coherent system with independent and heterogeneous component lifetimes, one can construct an equivalent coherent system with IID component lifetimes with the common reliability function \eqref{G-bar}.

\par \noindent As a suggested procedure for comparing various redundancy allocation policies, one may derive an equivalent system with IID component lifetimes  with the common reliability function \eqref{equivalent:G:bar} and then use the signature-based results for comparing systems. More precisely, assume one has  $k$ spars with lifetimes $Y_1,\cdots,Y_k$ and plan to allocate the spars to the $n$ (original) components. She has also two possible policies, say Policy I and Policy II. Under Policies I and II, the improved systems would have lifetimes $T^{[I]}$ and $T^{[II]}$ with signatures $\mathbf{s}^{[I]}=(s_1^{[I]},\cdots,s_{n+k}^{[I]})$ and $\mathbf{s}^{[II]}=(s_1^{[II]},\cdots,s_{n+k}^{[II]})$, respectively. Analogously to Equation \eqref{G-bar}, let 
\[\bar{G}^{[I]}(t)=h^{-1,[I]}\left(H^{[I]}(\bar{F}_1(t),\cdots,\bar{F}_{n+k}(t))\right),\ \ \ t>0,\]
and
\[\bar{G}^{[II]}(t)=h^{-1,[II]}\left(H^{[II]}(\bar{F}_1(t),\cdots,\bar{F}_{n+k}(t))\right),\ \ \ t>0,\]
where $\bar{F}_{n+1}(t),\cdots,\bar{F}_{n+k}(t)$ call for the reliability functions of the spare lifetimes $Y_1,\cdots,Y_k$. 
Here, $H^{[I]}$($H^{[II]}$) and $h^{[I]}$($h^{[II]}$) denote the structure system function and the reliability function of the improved system, respectively, under Policy I (Policy II).
Equation \eqref{equivalent:G:bar} yields for all $t>0$,
\begin{eqnarray}
\bar{F}_{T^{[I]}}(t)&=&H^{[I]}(\bar{F}_1(t),\cdots,\bar{F}_{n+k}(t))=\sum_{i=1}^{n+k} s_i^{[I]} \bar{G}^{[I]}_{i:n+k}(t),\label{PI:equivalent:G:bar}
\end{eqnarray}
and
\begin{eqnarray}
\bar{F}_{T^{[II]}}(t)&=&H^{[II]}(\bar{F}_1(t),\cdots,\bar{F}_{n+k}(t))=\sum_{i=1}^{n+k} s_i^{[II]} \bar{G}^{[II]}_{i:n+k}(t).\label{PII:equivalent:G:bar}
\end{eqnarray}
Now one can use the signature-based results for comparing the system lifetimes $T^{[I]}$ and $T^{[II]}$ with the reliability functions \eqref{PI:equivalent:G:bar} and \eqref{PII:equivalent:G:bar}, respectively. Here, some results which are useful in sequel are mentioned. For more information, see Chapter 4 of Samaniego \cite{samaniego2007}.
\begin{definition} Let $X$ and $Y$ be two random variables with reliability functions $\bar{F}$ and $\bar{G}$, respectively. Then $X$ is said to be smaller than $Y$ in stochastic order (hazard rate order, likelihood ration order), denoted by $X\leq_{st}(\leq_{hr},\leq_{lr})Y$ if $\bar{F}(t)\geq \bar{G}(t)$ for all $t$ ($\bar{G}(t)/\bar{F}(t)$, $\bar{g}(t)/\bar{f}(t)$ is increasing in $t$). Here, $f$ and $g$ are density functions of $X$ and $Y$, respectively.
\end{definition}
In the sequel and for all orderings above-mentioned, the statements ``$X\leq Y$" and ``$F\leq G$" are used interchangeably.
\begin{theorem}[Samaniego \cite{samaniego2007}, Chapter 4]\label{theo:order}
Let $\mathbf{s}_i$, $i=1,2$, denote the $i$-th system signature with IID component lifetimes and the common reliability function $\bar{G}_i$. 
\begin{enumerate}
\item If $\bar{G}_1(t)=\bar{G}_2(t)$ for $t>0$ and $\mathbf{s}_1\leq_{st,hr,lr}\mathbf{s}_2$ then $T_1\leq_{st,hr,lr}T_2$;
\item If $G_1\leq_{st}G_2$ and $\mathbf{s}_1=\mathbf{s}_2$ then $T_1\leq_{st}T_2$;
\end{enumerate}
\end{theorem}
\begin{example}\label{Example:2-component:ind:Active} Consider a 2-component series system and $k=1$ spare with two possible active redundancy policies $T^{[I]}=\min\{\max\{X_1,Y_1\},X_2\}$ and $T^{[II]}=\min\{X_1,\max\{X_2,Y_1\}\}$. Boland et al. \cite{Boland1992} proved that if the component lifetimes are independent and $X_1\leq_{st}X_2$ then $T^{[I]}\geq_{st} T^{[II]}$. Notice that in this case, $\mathbf{s}^{[I]}=\mathbf{s}^{[II]}=(1/3,2/3,0)$,
\begin{equation} \label{Exam1:RR2:HIand HII}
H^{[I]}(p_1,p_2,p_3)=(1-(1-p_1)(1-p_3))p_2,\ \  \ \ 
H^{[II]}(p_1,p_2,p_3)=p_1(1-(1-p_2)(1-p_3)).
\end{equation}
and hence $h^{[I]}(p)=h^{[II]}(p)=p(1-(1-p)^2)$ for $0<p<1$.
Moreover,
\begin{equation}\label{Exam1:PI:equivalent:G:bar}
\bar{F}_{T^{[I]}}(t)=H^{[I]}(\bar{F}_1(t),\bar{F}_{2}(t),\bar{F}_{3}(t))= (1-F_1(t)F_3(t))\bar{F}_2(t),
\end{equation}
and
\begin{equation}\label{Exam1:PII:equivalent:G:bar}
\bar{F}_{T^{[II]}}(t)=H^{[II]}(\bar{F}_1(t),\bar{F}_{2}(t),\bar{F}_{3}(t))=\bar{F}_1(t)(1-F_2(t)F_3(t)).
\end{equation}
The mathematical package MAPLE version 18 with procedure ``SOLVE'' gives
\begin{eqnarray}
h^{-1,[I]}(p)&=&h^{-1,[II]}(p)\nonumber\\
&=& \frac{1}{6}\,\sqrt [3]{64-108\,p+12\,\sqrt {81\,{p}^{2}-96\,p}}\nonumber\\
&&+\frac{8}{3}\,{\frac {1
}{\sqrt [3]{64-108\,p+12\,\sqrt {81\,{p}^{2}-96\,p}}}}+\frac{2}{3}, \ \ \forall\ 0<p<1.\label{Indep:h:inverse}
\end{eqnarray}
From Equations \eqref{G-bar} and \eqref{Exam1:RR2:HIand HII}, one can see that for all $t>0$,
\begin{eqnarray}
\bar{G}^{[I]}(t)&=&h^{-1,[I]}\left(H^{[I]}(\bar{F}_1(t),\cdots,\bar{F}_{n+k}(t))\right)\nonumber\\
&=&h^{-1,[I]}\left((1-F_1(t)F_3(t))\bar{F}_2(t)\right),\nonumber\\
\bar{G}^{[II]}(t)&=&h^{-1,[II]}\left(H^{[II]}(\bar{F}_1(t),\cdots,\bar{F}_{n+k}(t))\right)\nonumber\\
&=&h^{-1,[II]}\left(\bar{F}_1(t)(1-F_2(t)F_3(t))\right).\nonumber
\end{eqnarray}
Assume now that $X_1\leq_{st}X_2$ or $\bar{F}_1(t)\leq \bar{F}_2(t)$ for all $t$. Since $h^{-1,[I]}(p)=h^{-1,[II]}(p)$ are increasing in $p$ and after some algebraic calculations $(1-F_1(t)F_3(t))\bar{F}_2(t)\geq \bar{F}_1(t)(1-F_2(t)F_3(t))$ for all $t$, then $\bar{G}^{[I]}\geq_{st}\bar{G}^{[II]}$  and the above-mentioned result of Boland et al (1999) follows also by Part 2 of Theorem \ref{theo:order}. \hfill{$\Box$}
\end{example}

\begin{example}\label{Exam-Bridge}
Consider the five-component bridge system in Figure \ref{exam:Fig:Bridge}.
\begin{figure}[h!]
\includegraphics[scale=0.5]{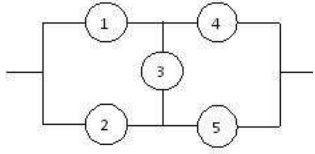}
\centering
\caption{The five-component bridge system.}
\label{exam:Fig:Bridge}
\end{figure}
We also have $k=1$ (active) spare and let $\mathbf{s}^{[i]}$, $i=1,\cdots,5$, denote the system signature when the spare has been redundant with the $i$-th component. One can verify that 
\[\mathbf{s}^{[1]}=\mathbf{s}^{[2]}=\mathbf{s}^{[4]}=\mathbf{s}^{[5]}=\left(0,\frac{1}{15},\frac{7}{30},\frac{1}{2},\frac{1}{5},0\right),\ \ \ \mathbf{s}^{[3]}=\left(0,\frac{2}{15},\frac{4}{15},\frac{7}{15},\frac{2}{15},0\right).\]
The reliability function of the bridge system (without the spare) is given by
\begin{eqnarray}
H(p_1,\cdots, p_5)&=&p_1 p_3+p_2 p_4+p_1 p_4 p_5
+p_2 p_3 p_5\nonumber\\
&&
-p_1 p_2 p_3 p_4-p_1 p_3 p_4 p_5
-p_1 p_2 p_3 p_5-p_2 p_3 p_4 p_5-p_1 p_2 p_4 p_5\nonumber\\
&&+2 p_1 p_2 p_3 p_4 p_5.\label{reliab:fun:bridge}
\end{eqnarray}
Then the reliability function of the system with an active spare which has been allocated to the $i$-th component, for $i=1,\cdots,5$, is derived from \eqref{reliab:fun:bridge} by replacing $p_i\cup p_6:=1-(1-p_i)(1-p_6)$ instead of $p_i$. Let $H^{[i]}(p_1,\cdots,p_6)$ denote the reliability function of the system when the active spare is allocated to the $i$-th component. Then from \eqref{G-bar}, the common reliability function of the equivalent system with IID components is derived as
\begin{equation}\label{G-bar-bridge}
\bar{G}^{[i]}(t)=h^{[i],-1}(H^{[i]}(\bar{F}_1(t),\cdots,\bar{F}_6(t))),\ \ \forall t>0,
\end{equation}
where $h^{[i]}(p)=H^{[i]}(p,\cdots,p)$. Using the mathematical package MAPLE version 18, we derived
\begin{eqnarray}
h^{[i]}(p)=-2\,{p}^{6}+8\,{p}^{5}-10\,{p}^{4}+2\,{p}^{3}+3\,{p}^{2}, \ \ \ i=1,\cdots,5.
\end{eqnarray}

Now let $X_i$, $i=1,\cdots,5$ be the independent exponentially distributed lifetimes with DF $F_i(t)=1-\exp\{-\lambda_i t\}$ and there is one active spare component ($k=1$) with the lifetime $X_6$. Similar to Boland et al. \cite{Boland1992}, the question is where to place the standby redundancy in order to make ``best'' improvement in the bridge system. The answer depends to the relative values of $\lambda_i$, $i=1,\cdots,5$ and the distribution of the spare lifetime. For illustration, let $F_6(t)=1-\exp\{-\lambda_6 t\}$ for $t>0$. Table \ref{tab:optimal:bridge:system} displays the optimal allocation for some selected values of $\lambda_i$, $i=1,\cdots,6$. In some cases, one can not determine the optimal policy and another reliability index may used such as the mean time to failure (MTTF) of the improved system. For example, let $W(\alpha,\lambda)$ stand for the Weibull distribution with DF 
\[F(t)=1-\exp\{-(\lambda t)^\alpha\},\ \ \ t>0, \ \lambda>0,\ \ \alpha>0.\] 
Figure \ref{exam:Fig:Bridge:weibull} pictures the reliability function of the bridge system with one spare when the component lifetime $X_i$ follows the Weibull distribution $W(\alpha_i,1)$ for $i=1,\cdots,6$. Here, $\alpha_1=\alpha_4=\alpha_6=2$, $\alpha_2=\alpha_5=0.5$ and $\alpha_3=1$.
\begin{figure}[h!]
\includegraphics[scale=0.5]{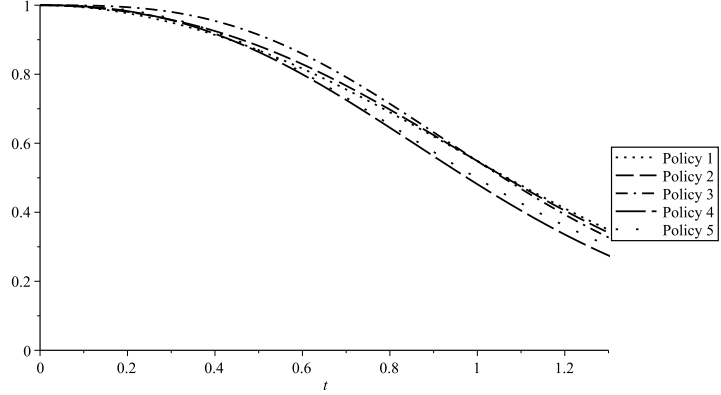}
\centering
\caption{The reliability function of the bridge system under various redundancy allocations.}
\label{exam:Fig:Bridge:weibull}
\end{figure}
The calculations have been done by the mathematical package MAPLE version 18. The corresponding program is given in the appendix.
\begin{table}
\label{tab:optimal:bridge:system}
\centering
\caption{The optimal allocation for the bridge system in Figure \ref{exam:Fig:Bridge}.}
\begin{tabular}{ccccccc}\hline
$\lambda_1$&$\lambda_2$&$\lambda_3$&$\lambda_4$&$\lambda_5$&$\lambda_6$&optimal component for allocation\\ \hline
1&1&1&2&2&1.5&4\\
1&1&1.5&2&2&1.5&4\\
1&1&2&2&2&1.5&4\\ \\

1&1&1&2&2&1&4\\
1&1&1&2&2&1.5&4\\
1&1&1&2&2&2&4\\ \\

1&2&1  &1&2&1.5&2\\
1&2&1.5&1&2&1.5&2\\
1&2&2  &1&2&1.5&2\\ \\

2&1&1&1&2&1.5&2\\
2&1&1&2&1&1.5&1 or 4\\
3&1&1&2&1&1.5&1\\ \hline
\end{tabular}
\end{table}
\hfill{$\Box$}
\end{example}

The proposed signature-based approach may be applied also to standby redundancy policies. The next example illustrates this approach. In the example and hereafter, $F*G(t)$ means convolution of two DFs $F$ and $G$, defined by
\[F*G(t)=\int_{-\infty}^t G(t-x)d F(x),\]
and $\overline{F*G}(t)=1-F*G(t)$ for all $t$.
\begin{example}\label{Exam2}
Consider again the 2-component series system and $k=1$ spare with two possible standby redundancy policies $T^{[I]}=\min\{X_1+Y_1,X_2\}$ and $T^{[II]}=\min\{X_1,X_2+Y_1\}$. 
Theorem 3.2 of Boland et al. \cite{Boland1992} states that if the component lifetimes are independent and $X_1\leq_{st}X_2$ then $T^{[I]}\geq_{st} T^{[II]}$ provided that the component lifetimes possess ``the reverse rule of order 2 property"; For more information, see the appendix and also Karlin \cite{Karlin}. Notice that in this case one has two two-component series systems with the component reliability functions $(\overline{F_1*F_3},\bar{F_2})$ and $(\bar{F_1},\overline{F_2*F_3})$ under Policies I and II, respectively. So $\mathbf{s}^{[I]}=\mathbf{s}^{[II]}=(1,0)$, $H^{[I]}(p_1,p_2)=H^{[II]}(p_1,p_2)=p_1 p_2$, $h^{[I]}(p)=h^{[II]}(p)=p^2$ and then $h^{-1,[I]}(p)=h^{-1,[II]}(p)=\sqrt{p}$.
Equation \eqref{equivalent:G:bar} gives
\begin{eqnarray}
\bar{F}_{T^{[I]}}(t)&=&H^{[I]}(\overline{F_1*F_3}(t),\bar{F_2}(t))=\overline{F_1*F_3}(t)\bar{F_2}(t),\ \ \ t>0,\label{Exam2:PI:equivalent:G:bar}\\
\bar{F}_{T^{[II]}}(t)&=&H^{[II]}(\bar{F_1}(t),\overline{F_2*F_3}(t))=\bar{F_1}(t)\overline{F_2*F_3}(t),\ \ \ t>0.
\label{Exam2:PII:equivalent:G:bar}
\end{eqnarray}
Equation \eqref{G-bar} yields
\begin{eqnarray}
\bar{G}^{[I]}(t)&=&\sqrt{\overline{F_1*F_3}(t)\bar{F_2}(t)},\label{Exam2:PI:equivalent:G:bar}\\
\bar{G}^{[II]}(t)&=&\sqrt{\bar{F_1}(t)\overline{F_2*F_3}(t)}.\label{Exam2:PII:equivalent:G:bar}
\end{eqnarray}
Therefore, for comparison purposes, one just needs to consider the reliability functions given by Equations \eqref{Exam2:PI:equivalent:G:bar} and \eqref{Exam2:PII:equivalent:G:bar} since both systems have identical system signatures. Then, if $\bar{G}^{[I]}(t)\geq \bar{G}^{[II]}(t)$, for all $t>0$, then $T^{[I]}\geq_{st} T^{[II]}$ by Part 2 of Theorem \ref{theo:order}.

For example, suppose
$X_1$ follows the exponential distribution with the DF $F_2(t)=1-\exp\{-t\}$, $t>0$ and $X_2$ has the Pareto distribution with the DF $F_1(t)=1-(1+t)^{-1}$, $t>0$. Moreover, the spare lifetime $Y_1\sim F_3(t)=1-\exp\{-2t\},$ for $t>0$.
Using the mathematical software MAPLE version 18, the graphs of $\bar{G}^{[I]}(t)$ and $\bar{G}^{[II]}(t)$, given by Equations \eqref{Exam2:PI:equivalent:G:bar} and \eqref{Exam2:PII:equivalent:G:bar}, are pictured in Figure \ref{exam:Fig}.

\begin{figure}[h!]
\includegraphics[scale=0.3]{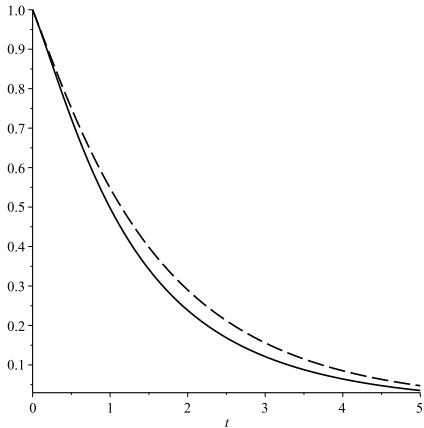}
\centering
\caption{Graphs of $\bar{G}^{[I]}(t)$ (Dashed line) and $\bar{G}^{[II]}(t)$ (Solid line) in Example \ref{Exam2}.}
\label{exam:Fig}
\end{figure}
As one can see from Figure \ref{exam:Fig}, $G^{[I]}\geq_{st}G^{[II]}$ then $T^{[I]}\geq_{st} T^{[II]}$ by Part 2 of Theorem \ref{theo:order}. Note that in this example the family $\{F_1,F_2\}$ does not posses the RR2 property. To see this, the respective densities are $f_1(t)=(1+t)^{-2}$ and $f_2(t)=\exp\{-t\}$, for $t>0$. Let $x_1=1$ and $x_2=2$. One can easily see that $f_1(x_1)f_2(x_2)=0.0338\ngeq f_1(x_2)f_2(x_1)=0.04088$. Hence, Theorem 3.2 of Boland et al. \cite{Boland1992} can not be applied in this RAP.
\hfill{$\Box$}
\end{example}
\section{RAP with dependent component lifetimes}
In practice, the system components may share the same environmental factors such as temperature, pressure, loading and etc. Then,  the component lifetimes are not independent, but rather are ``associated" and exhibit some dependency. 
Examples include structures in which components share the load, so that failure of one component results in increased load on each of the remaining components. For more information, see Barlow and Proschan \cite{BP1975} and Nelsen \cite{nelson2006}. The RAP for systems with dependent component lifetimes has not been extensively studied in literature. Among few works, Kotz et al. \cite{kotz2003} investigated the increase in the mean lifetime for parallel redundancy when the two component lifetimes are positive (negative) dependent.
da Costa Bueno \cite{bueno2005} defined the concept of ``minimal standby redundancy" and used the reverse rule of order 2 property between component lifetimes to study the problem of RAP for $k$-out-of-$n$ systems with dependent components using a martingale approach. See also da Costa Bueno and Martins do Carmo \cite{Iran}.
 Belzunce \textit{et al.} \cite{belzunce2011,belzunce2013} considered  the RAP \rd{with dependent} component lifetimes. For modelling the dependency among  component lifetimes and comparison purposes, they used the concept of ``joint stochastic orders''. You and  Li \cite{youli2014}  extended the result of Boland et al. \cite{Boland1992} from independent components  to allocating \textit{m} independent and identically distributed (IID) active redundancy lifetimes to $k$-out-of-$n$ system with components lifetimes having an arrangement increasing (AI)
joint density. They proved that  more redundancies should be allocated to the weaker component to increase the reliability of the system. Recently, Jeddi and Doostparast \cite{jeddi2015} considered the RAP without any restriction to a special structure \rd{form} for dependency among component lifetimes. Their conditions are \rd{expressed} in terms of the joint distribution of the component lifetimes. This section deals with the problem of allocating spare components via the proposed 
signature-based approach for improving the system reliability in which the lifetimes of components are dependent.\\

Navarro et al. \cite{Navarro2011} extended th representation \eqref{equivalent:G:bar} based on signatures to coherent systems with component lifetimes that may be dependent. To describe the results, let $T=\phi(X_1,\cdots,X_n)$ be the lifetime of a coherent system with structure function $\phi$ and component lifetimes $X_1,\cdots,X_n$ with the joint reliability function $\bar{F}_{X_1,\cdots,X_n}(x_1,\cdots,x_n)=P(X_1>x_1,\cdots,X_n>x_n)$.
Sklar's theorem (Nelsen, \cite{nelson2006}, p. 46) ensures
\begin{equation}\label{joint:survival}
\bar{F}_{X_1,\cdots,X_n}(x_1,\cdots,x_n)=K\left(\bar{F}_{X_1}(x_1),\cdots,\bar{F}_{X_n}(x_n)\right),
\end{equation}
where $\bar{F}_{X_i}(x_i)=P(X_i>x_i)$, for $i=1,\cdots,n$, is the marginal reliability function of component lifetime $X_i$ and $K$ is the survival copula. The coherent system lifetime $T$ may be represented as $T=\max_{1\leq j\leq l}X_{P_j}$ where $X_{P_j}=\min_{i\in P_j}X_i$ and $P_1,\cdots,P_l$ stand for the all ``minimal paths'' of the system. For more information, see Barlow and Proschan \cite{BP1975}. Hence,  the system reliability function  can be written as (Navarro et al. \cite{Navarro2011})
\begin{equation}\label{Depend:W:Survival}
\bar{F}_T(t)=W\left(\bar{F}_{X_1}(x_1),\cdots,\bar{F}_{X_n}(x_n)\right),
\end{equation}
where 
\[W(x_1,\cdots,x_n)=\sum_{j=1}^l K(\mathbf{x}_{P_j})-\sum_{i<j} K(\mathbf{x}_{P_i \bigcup P_j})+\cdots+(-1)^{l+1}K(\mathbf{x}_{P_1\bigcup\cdots \bigcup P_l}),\]
and $\mathbf{x}_P=(z_1,\cdots,z_n)$ with $z_i=x_i$ for $i\in P$ and $z_i=1$ for $i\notin P$. Here $W=W(\phi,K)$ is known as ``structure-dependence function ''. In particular, if the component lifetimes are independent then the function $W$ is equal to the structure reliability function $H$ in Equation \eqref{equivalent:G:bar}. 
\begin{example}[Navarro et al. \cite{Navarro2011}]\label{Depend:exam1}
Consider a system with lifetime $T=\min(X_1,\max(X_2,X_3))$. Then $P_1=\{1,2\}$ and $P_2=\{1,3\}$. By Equation \eqref{Depend:W:Survival}
\begin{equation}
\bar{F}_T(t)=W(\bar{F}_1(t),\bar{F}_2(t),\bar{F}_3(t)),
\end{equation}
where $W(x_1,x_2,x_3)=K(x_1,x_2,1)+K(x_1,1,x_3)-K(x_1,x_2,x_3)$.
Notice that if the component lifetimes are independent, then $K(x_1,x_2,x_3)=x_1 x_2 x_3$ and $W(x_1,x_2,x_3)=x_1 x_2+x_1 x_3-x_1 x_2 x_3$.\\

\hfill{$\Box$}
\end{example}

Navarro et al. \cite{Navarro2011} proved that the system lifetime $T$ is equal in law with $T^\star=\phi(Y_1,\cdots,Y_n)$ where $Y_1,\cdots,Y_n$ are identically distributed with the joint  reliability function 
\begin{equation}\label{Depen:joint:survival:of:Yi}
P(Y_1>x_1,\cdots,Y_n>x_n)=K(\bar{G}_W(x_1),\cdots,\bar{G}_W(x_n)),
\end{equation}
where $\bar{G}_W(t)=m_W(\bar{F}_1(t),\cdots,\bar{F}_n(t))$ and $m_W(x_1,\cdots,x_n)$ is ``the  mean function'' of $W$, defined by
$m_W(x_1,\cdots,x_n)=\delta^{-1}(W(x_1,\cdots,x_n))$ on the space $[0,1]^n$ and $\delta(x)=W(x,\cdots,x)$ for $x\in [0,1]$.
Moreover, if the survival copula $K$ is exchangeable, then
\begin{equation}
\bar{F}_T(t)=\sum_{i=1}^n s_i \bar{G}_{i:n}(t),
\end{equation}
where $\bar{G}_{i:n}(t)=P(Y_{i:n}>t)$ and $Y_{1:n}<\cdots<Y_{n:n}$ are the order statistics obtained from the random variables $Y_1,\cdots,Y_n$ with the joint reliability function \eqref{Depen:joint:survival:of:Yi}. 
The next theorem is valuable in RAP for coherent systems.
\begin{theorem}[Navarro et al. \cite{Navarro2011}]\label{Depen:theo:equivalence}
If $T$ is the lifetime of a coherent system with signature $\mathbf{s}=(s_1,\cdots,s_n)$ and with component lifetimes $X_1,\cdots,X_n$ having the structure-dependence function $W$, then 
\begin{equation}\label{Depen:survival:equivalence}
\bar{F}_T(t)=\sum_{i=1}^n s_i \bar{G}_{i:n}(t),
\end{equation}
where   $\bar{G}_{i:n}(t)=P(Y_{i:n}>t)$ and $Y_{1:n}<\cdots<Y_{n:n}$ stand for the order statistics obtained from the IID random variables $Y_1,\cdots,Y_n$ with the common reliability function $\bar{G}(t)=h^{-1}(W(\bar{F}_{X_1}(t),\cdots,\bar{F}_{X_n}(t)))$ and $h$ is the reliability polynomial of the coherent system.
\end{theorem}

\begin{example}\label{exam:mardia}
Suppose that the component lifetimes in Example \ref{Example:2-component:ind:Active} are dependent 
and follow the Mardia tri-variate Pareto distribution with the joint reliability function (Mardia \cite{Mardia1962})
\begin{equation}
\label{pdf:Mardia}
\bar{F}_{X_1,X_2,X_3}(x_1,x_2,x_3)=\left(\frac{x_1}{\sigma_1}+\frac{x_2}{\sigma_2}+\frac{x_3}{\sigma_3}-2\right)^{-\alpha},\ \ x_i>\sigma_i>0,\ \ i=1,2,3,\ \ \alpha>0.
\end{equation}
Jeddi and Doostparast \cite{jeddi2015} proved that $T^{[I]}\geq_{st}T^{[II]}$ provided $X_1\leq_{st}X_2$. Here a signature-based approach is utilized. To do this note that similar to Example \ref{Depend:exam1}, the reliability system functions under Policies I and II, respectively, are given by
$\bar{F}_{T^{[I]}}(t)=W^{[I]}(\bar{F}_1(t),\bar{F}_2(t),\bar{F}_3(t))$ and $\bar{F}_{T^{[II]}}(t)=W^{[II]}(\bar{F}_1(t),\bar{F}_2(t),\bar{F}_3(t))$, where
\begin{eqnarray}
W^{[I]}(x_1,x_2,x_3)&=&K(x_1,x_2,1)+K(1,x_2,x_3)-K(x_1,x_2,x_3),\label{dep:exam:1:WI}\\
W^{[II]}(x_1,x_2,x_3)&=&K(x_1,x_2,1)+K(x_1,1,x_3)-K(x_1,x_2,x_3),
\end{eqnarray}
with the survival copula $K(x_1,x_2,x_3)=\left(x_1^{-1/\alpha}+
x_2^{-1/\alpha}+x_3^{-1/\alpha}-2\right)^{-\alpha}$,
which known as ``Clyton copula''; See Nelsen \cite{nelson2006} for more information. Notice that the popular form for the Clyton copula is as follow:
\begin{equation}
\label{Clyton-copula}
K(x_1,x_2,x_3)=\left(x_1^{-\theta}+x_2^{-\theta}+x_3^{-\theta}-2\right)^{-1/\theta},\ \ \theta\geq 0.
\end{equation} 
By Theorem \ref{Depen:theo:equivalence}, the equivalent systems under Policies I and II with IID component lifetimes have the common component reliability functions
\begin{eqnarray}
\bar{G}^{[I]}(t)&=&h^{-1,[I]}\left(W^{[I]}(\bar{F}_1(t),\bar{F}_2(t),\bar{F}_3(t))\right)\nonumber\\
&=&h^{-1,[I]}\left(K(\bar{F}_1(t),\bar{F}_2(t),1)+K(1,\bar{F}_2(t),\bar{F}_3(t))-K(\bar{F}_1(t),\bar{F}_2(t),\bar{F}_3(t))\right),\label{dep:G:bar:I}\\
\bar{G}^{[II]}(t)&=&h^{-1,[II]}\left(W^{[II]}(\bar{F}_1(t),\bar{F}_2(t),\bar{F}_3(t))\right)\nonumber\\
&=&h^{-1,[II]}\left(K(\bar{F}_1(t),\bar{F}_2(t),1)+K(\bar{F}_1(t),1,\bar{F}_3(t))-K(\bar{F}_1(t),\bar{F}_2(t),\bar{F}_3(t))\right),\label{dep:G:bar:II}\end{eqnarray}
respectively, where the functions $h^{-1,[I]}(p)$ and $h^{-1,[II]}(p)$ are given by Equation \eqref{Indep:h:inverse}. 
Since $\mathbf{s}^{[I]}=\mathbf{s}^{[II]}$, one solely needs to compare the component reliability functions \eqref{dep:G:bar:I} and \eqref{dep:G:bar:II} and then uses Part 2 of Theorem \ref{theo:order}. Therefore, $T^{[I]}\geq_{st}T^{[II]}$ if and only if $\bar{G}^{[I]}(t)\geq \bar{G}^{[II]}(t)$ for all $t>0$. 
Notice that the functions $h^{-1,[I]}(p)=h^{-1,[II]}(p)$ are increasing in $p$. 
 From \eqref{dep:exam:1:WI}-\eqref{dep:G:bar:II} and after simple algebraic calculations, one can see that $\bar{G}^{[I]}(t)\geq \bar{G}^{[II]}(t)$ for all $t>0$ holds if and only if $\bar{F}_1(t)\leq \bar{F}_2(t)$ for all $t>0$. Hence, $T^{[I]}\geq_{st}T^{[II]}$ if and only if $X_1\leq_{st}X_2$ as proved by Jeddi and Doostparast \cite{jeddi2015}.
\hfill{$\Box$}
\end{example}

\noindent Here, we assumed that the dependence structure does not change under different policies. If the dependency structure change under various policies, the above-mentioned signature-based procedure may be used to derive an optimal RAP. For example assume that the dependency among component lifetimes under Policy I follows the Clyton copula \eqref{Clyton-copula} while under Policy II follows ``the Gumbel copula'' 
\begin{equation}\label{Gumbel-copula}
K(x_1,x_2,x_3)=\exp\left(\left\{(-\log x_1)^\gamma+(-\log x_2)^\gamma+(-\log x_3)^\gamma\right\}^{1/\gamma}\right),\ \ x_i>0,\ \ i=1,2,3.
\end{equation}
where ``$\log$'' stands for the natural logarithm. Moreover, let $\bar{F}_i(t)=(1+t)^i$, $t>0$ for $i=1,2,3$, the Pareto distribution. Upon substituting Equations \eqref{Clyton-copula} and \eqref{Gumbel-copula}, respectively, into Equations \eqref{dep:G:bar:I} and \eqref{dep:G:bar:II}, the component reliability functions $\bar{G}^{[I]}(t)$ and $\bar{G}^{[II]}(t)$ are derived. A graph of these functions may be useful to derive the optimal policy. 
\section{Conclusion}
This paper dealt with a signature-based approach for redundancy allocation problems when component lifetimes are either heterogeneous or dependent. An important part to implement the derived results is derivation of the system signature.  There are some researches to obtain system signatures; See, e.g., Gertsbakh et al. \cite{Gertsbakh}, Marichal and Mathonet \cite{Marichal} and Navarro and Rubino \cite{Navarro2010} and references therein. 
\section*{Acknowledgement}
This research was supported by a grant from Ferdowsi University of Mashhad (No MS94329DSP).

\section*{Appendix}
\subsection*{The reverse rule of order 2 property}
\begin{definition}
The function $g(\theta,x)$ has the reverse rule of order 2 (denoted by RR2) property in $\theta$ and $x$ if for $\theta_1>\theta_2$ and $x_1<x_2$,
\[g(\theta_1,x_1)g(\theta_2,x_2)\geq g(\theta_1,x_2)g(\theta_2,x_1).\]
\end{definition}
Many one-parameter families of density functions $\{g(\theta,x):=f_\theta(x), \theta\in\Theta\}$ of life distributions possess the RR2 property. Examples include:
\begin{itemize}
\item Gamma distribution $\Gamma(m,\lambda)$ with density \[f_{m,\lambda}(x)=\frac{\lambda^m}{\Gamma(m)}x^{m-1}\exp\{-\lambda x\},\ \ \ x>0.\]
The families $\{\Gamma(m,\theta),\theta>0\}$ (fixed shape parameter) and $\{\Gamma(\theta^{-1},\lambda),\theta>0\}$ (fixed scale parameter);
\item Weibull distribution with density
$f_\theta(x)=\alpha\theta x^{\alpha-1}\exp\{-\theta x^\alpha\}$, for $\lambda>0$, is RR2 with fixed shape parameter $\alpha$;
\item Pareto family of densities $\{f_\theta(x)=\theta (1+x)^{-(\theta+1)},\theta>0\}$.
\end{itemize}
\subsection*{MAPLE codes for the bridge example}
\begin{itemize}
\item H(p1,p2,p3,p4, p5):=p1* p3+p2* p4+p1* p4* p5+p2* p3* p5 -p1* p2* p3* p4-p1* p3* p4* p5-p1* p2* p3* p5-p2* p3* p4* p5-p1* p2* p4* p5 +2 p1* p2* p3* p4* p5); h(p):=H(p,p,p,p,p);

\item H1(p1,p2,p3,p4, p5,p6):=H(1-(1- p1)*(1-p6),p2,p3,p4, p5);  h1(p):=H1(p,p,p,p, p,p); 

\item H2(p1,p2,p3,p4, p5,p6):=H(p1,1-(1- p2)*(1-p6),p3,p4, p5);  h2(p):=H2(p,p,p,p, p,p) ; 

\item H3(p1,p2,p3,p4, p5,p6):=H(p1,p2,1-(1- p3)*(1-p6),p4, p5);  h3(p):=H3(p,p,p,p, p,p) ; 

\item H4(p1,p2,p3,p4, p5,p6):=H(p1,p2,p3,1-(1- p4)*(1-p6), p5);  h4(p):=H4(p,p,p,p, p,p);  

\item H5(p1,p2,p3,p4, p5,p6):=H(p1,p2,p3,p4,1-(1- p5)*(1-p6));  h5(p):=H5(p,p,p,p, p,p); 
\item simplify(h1(p)); simplify(h2(p)); simplify(h3(p)); simplify(h4(p)); simplify(h5(p));

\item f := x $\rightarrow -2*x^6+8*x^5-10*x^4+2*x^3+3*x^2$;\\
solve(x = f(y), y) assuming 0 $\leq$ y $\leq$ 1 and 0 $\leq$ x $\leq$ 1; \\
g := unapply($\%$, x); g(f(x));
\item F1bar := t$\rightarrow\exp\{-(\lambda_1 t)^{\alpha_1}\}$;\\
F2bar := t$\rightarrow\exp\{-(\lambda_2 t)^{\alpha_2}\}$;\\
F3bar := t$\rightarrow\exp\{-(\lambda_3 t)^{\alpha_3}\}$;\\
F4bar := t$\rightarrow\exp\{-(\lambda_4 t)^{\alpha_4}\}$;\\
F5bar := t$\rightarrow\exp\{-(\lambda_5 t)^{\alpha_5}\}$;\\
F6bar := t$\rightarrow\exp\{-(\lambda_6 t)^{\alpha_6}\}$;\item n := 5; s1245 := (0, 1/15, 7/30, 1/2, 1/5, 0); s3 :=( 0, 2/15, 4/15, 7/15, 2/15, 0); $\lambda_1 := 1$; $\lambda_2:= 1$; $\lambda_3:= 1$; $\lambda_4:= 1$; $\lambda_5:= 1$; $\lambda_6:= 1$;\\
$\alpha_1:= 2$; $\alpha_2:= 0.5$; $\alpha_3:= 1$; $\alpha_4:= 2$; $\alpha_5:= 0.5$; $\alpha_6:= 2$;
\item E(n,i,a):=$\sum_{j=i}^n  a^{j}(1-a)^{n-j}$
\item Gbar1nPolicy1 := 1-E(n, 1, 1-g(H1(F1bar(t), F2bar(t), F3bar(t), F4bar(t), F5bar(t), F6bar(t)))); Gbar2nPolicy1 := 1-E(n, 2, 1-g(H1(F1bar(t), F2bar(t), F3bar(t), F4bar(t), F5bar(t), F6bar(t)))); Gbar3nPolicy1 := 1-E(n, 3, 1-g(H1(F1bar(t), F2bar(t), F3bar(t), F4bar(t), F5bar(t), F6bar(t)))); Gbar4nPolicy1 := 1-E(n, 4, 1-g(H1(F1bar(t), F2bar(t), F3bar(t), F4bar(t), F5bar(t), F6bar(t)))); Gbar5nPolicy1 := 1-E(n, 5, 1-g(H1(F1bar(t), F2bar(t), F3bar(t), F4bar(t), F5bar(t), F6bar(t)))); Gbar1nPolicy2 := 1-E(n, 1, 1-g(H2(F1bar(t), F2bar(t), F3bar(t), F4bar(t), F5bar(t), F6bar(t)))); Gbar2nPolicy2 := 1-E(n, 2, 1-g(H2(F1bar(t), F2bar(t), F3bar(t), F4bar(t), F5bar(t), F6bar(t)))); Gbar3nPolicy2 := 1-E(n, 3, 1-g(H2(F1bar(t), F2bar(t), F3bar(t), F4bar(t), F5bar(t), F6bar(t)))); Gbar4nPolicy2 := 1-E(n, 4, 1-g(H2(F1bar(t), F2bar(t), F3bar(t), F4bar(t), F5bar(t), F6bar(t)))); Gbar5nPolicy2 := 1-E(n, 5, 1-g(H2(F1bar(t), F2bar(t), F3bar(t), F4bar(t), F5bar(t), F6bar(t)))); Gbar1nPolicy3 := 1-E(n, 1, 1-g(H3(F1bar(t), F2bar(t), F3bar(t), F4bar(t), F5bar(t), F6bar(t)))); Gbar2nPolicy3 := 1-E(n, 2, 1-g(H3(F1bar(t), F2bar(t), F3bar(t), F4bar(t), F5bar(t), F6bar(t)))); Gbar3nPolicy3 := 1-E(n, 3, 1-g(H3(F1bar(t), F2bar(t), F3bar(t), F4bar(t), F5bar(t), F6bar(t)))); Gbar4nPolicy3 := 1-E(n, 4, 1-g(H3(F1bar(t), F2bar(t), F3bar(t), F4bar(t), F5bar(t), F6bar(t)))); Gbar5nPolicy3 := 1-E(n, 5, 1-g(H3(F1bar(t), F2bar(t), F3bar(t), F4bar(t), F5bar(t), F6bar(t)))); Gbar1nPolicy4 := 1-E(n, 1, 1-g(H4(F1bar(t), F2bar(t), F3bar(t), F4bar(t), F5bar(t), F6bar(t)))); Gbar2nPolicy4 := 1-E(n, 2, 1-g(H4(F1bar(t), F2bar(t), F3bar(t), F4bar(t), F5bar(t), F6bar(t)))); Gbar3nPolicy4 := 1-E(n, 3, 1-g(H4(F1bar(t), F2bar(t), F3bar(t), F4bar(t), F5bar(t), F6bar(t)))); Gbar4nPolicy4 := 1-E(n, 4, 1-g(H4(F1bar(t), F2bar(t), F3bar(t), F4bar(t), F5bar(t), F6bar(t)))); Gbar5nPolicy4 := 1-E(n, 5, 1-g(H4(F1bar(t), F2bar(t), F3bar(t), F4bar(t), F5bar(t), F6bar(t)))); Gbar1nPolicy5 := 1-E(n, 1, 1-g(H5(F1bar(t), F2bar(t), F3bar(t), F4bar(t), F5bar(t), F6bar(t)))); Gbar2nPolicy5 := 1-E(n, 2, 1-g(H5(F1bar(t), F2bar(t), F3bar(t), F4bar(t), F5bar(t), F6bar(t)))); Gbar3nPolicy5 := 1-E(n, 3, 1-g(H5(F1bar(t), F2bar(t), F3bar(t), F4bar(t), F5bar(t), F6bar(t)))); Gbar4nPolicy5 := 1-E(n, 4, 1-g(H5(F1bar(t), F2bar(t), F3bar(t), F4bar(t), F5bar(t), F6bar(t)))); Gbar5nPolicy5 := 1-E(n, 5, 1-g(H5(F1bar(t), F2bar(t), F3bar(t), F4bar(t), F5bar(t), F6bar(t))))

\item FbarSystemPolicy1 := s1245[1]*Gbar1nPolicy1(t)+s1245[2]*Gbar2nPolicy1(t)\\+s1245[3]*Gbar3nPolicy1(t)+s1245[4]*Gbar4nPolicy1(t)+s1245[5]*Gbar5nPolicy1(t);\\
 FbarSystemPolicy2 := s1245[1]*Gbar1nPolicy2(t)+s1245[2]*Gbar2nPolicy2(t)\\+s1245[3]*Gbar3nPolicy2(t)+s1245[4]*Gbar4nPolicy2(t)+s1245[5]*Gbar5nPolicy2(t); \\FbarSystemPolicy3 := s3[1]*Gbar1nPolicy3(t)+s3[2]*Gbar2nPolicy3(t)\\+s3[3]*Gbar3nPolicy3(t)+s3[4]*Gbar4nPolicy3(t)+s3[5]*Gbar5nPolicy3(t); \\FbarSystemPolicy4 := s1245[1]*Gbar1nPolicy4(t)+s1245[2]*Gbar2nPolicy4(t)\\+s1245[3]*Gbar3nPolicy4(t)+s1245[4]*Gbar4nPolicy4(t)+s1245[5]*Gbar5nPolicy4(t); \\FbarSystemPolicy5 := s1245[1]*Gbar1nPolicy5(t)+s1245[2]*Gbar2nPolicy5(t)\\
 +s1245[3]*Gbar3nPolicy5(t)+s1245[4]*Gbar4nPolicy5(t)+s1245[5]*Gbar5nPolicy5(t)

\item plot({FbarSystemPolicy1, FbarSystemPolicy2, FbarSystemPolicy3, FbarSystemPolicy4, FbarSystemPolicy5}, t = 0 .. 1.5)
\end{itemize}


\begin{thebibliography}{99}
\bibitem{BP1975}
\rd{Barlow, R. E., Proschan, F. (1975).
\newblock {\textit{Statistical theory of reliability and life testing}.}
\newblock { Holt, Rinehart and Winston, Inc.}, New York.}

\bibitem{belzunce2011}
Belzunce, F., Martinez-Puertas, H. Ruiz, J.M. (2011).
\newblock {On optimal allocation of redundant components for series and parallel systems of two dependent components,}
\newblock {\em Journal of Statistical Planning and Inference,} \textbf{141}, 3094-3104.

\bibitem{belzunce2013}
Belzunce, F., Martinez-Puertas, H. Ruiz, J.M. (2013).
\newblock {On allocation of redundant components for systems with dependent components,}
\newblock {\em European Journal of Operational Research,} \textbf{230}, 573-580.

\bibitem{Boland1992}
Boland, P.J.,~ EI-Neweihi, E.,~ Porschan, F.  (1992).
\newblock { Stochastic order for redundancy allocations in series and parallel systems},
\newblock {\em Advances in Applied Probability, } \textbf{24}, 161-171.

\bibitem{Boland1992}
Boland, P.J.,~ EI-Neweihi, E.,~ Porschan, F.  (1992).
\newblock { Stochastic order for redundancy allocations in series and parallel systems},
\newblock {\em Advances in Applied Probability, } \textbf{24}, 161-171.

\bibitem{bueno2005}
\rd{da Costa Bueno, V., (2005).
\newblock {Minimal standby redundancy allocation in a $k$-out-of-$n$:$F$ system of dependent components}.
\newblock {\em European Journal of Operational Research,} \textbf{165}, 786-793.}

\bibitem{Iran}
\rd{da Costa Bueno, V., Martins do Carmo, I. (2007).
\newblock {Active redundancy allocation for a $k$-out-of-$n$:$F$ system of dependent components}.
\newblock {\em European Journal of Operational Research,} \textbf{176}, 1041-1051.}

\bibitem{Gertsbakh}
Gertsbakh, I., Shpungin, Y. and Spizzichino, F. (2011).
\newblock { Signatures of coherent systems built with separate modules}.
\newblock {\em Journal of Applied Probability,} \textbf{48}, 843-855.

\bibitem{hu2008}
Hu, T., Wang, Y. (2009).
\newblock {Optimal allocation of active redundancies in r-out-of-n systems}.
\newblock {\em Journal of Statistical Planning and Inference,} \textbf{139}, 3733-3737.


\bibitem{jeddi2015}
Jeddi, M., Doostparast, M. (2016).
\newblock {Optimal redundancy allocation problems in engineering systems with dependent component lifetimes}.
\newblock {\em Applied Stochastic Models in Business and Industry,} \textbf{32}, 199-208.



\bibitem{Karlin} 
Karlin, S. (1968).
\newblock{\em Total Positivity}.
\newblock {Stanford University Press.}


\bibitem{kotz2003}
Kotz, S., Lai, C. D., Xie, M., (2003).
\newblock {On the effect of redundancy for systems with dependent components}.
\newblock {\em IIE Transactions,} \textbf{35}, 1103-1110.




\bibitem{Lehmann1966} Lehmann, E. L. (1966). 
\newblock {Some concepts of dependence}.
\newblock {\em The Annals of Mathematical Statistics,} \textbf{37}, 1137-1153. 



\bibitem{Mardia1962}
Mardia, K. V. (1962).
\newblock {Multivariate Pareto distributions}.
\newblock {\em The Annals of Mathematical Statistics} \textbf{33}, 1008-1015. 

\bibitem{Marichal}
Marichal, J.-L. and Mathonet, P. (2013).
\newblock {Computing system signatures through reliability functions}.
\newblock {\em Statistics and Probability Letters} \textbf{83}, 710-717. 

\bibitem{meeker11998}
Meeker, W. Q., Escobar, L. A. (1998).
\newblock{\em Statistical Methods for Reliability Data}.
\newblock { John Wiley \& Sons, New York.}



\bibitem{mi1999}
\rd{Mi, J. (1999).
\newblock {Optimal active redundancy,} 
\newblock {\em Journal of Applied Probability,} \textbf{31}, 1004-1014.}

\bibitem{Navarro2010}
\rd{Navarro, J., and Rubino, R. (2010).}
\newblock {Computations of Signatures of Coherent Systems with Five Components}.
\newblock {\em Communications in Statistics-Simulation and Computation,} \textbf{39},  68-84.


\bibitem{Navarro2011}
\rd{Navarro, J., Samaniego, F. J., Balakrishnan, N., (2011).}
\newblock {Signature-based representations for the reliability of systems with heterogeneous components}.
\newblock {\em Journal of Applied Probability,} \textbf{48},  856-867.

\bibitem{nelson2006}
\rd{Nelsen, R. B. (2006).}
\newblock {\textit{An introduction to copulas}.}
\newblock { Springer}, New York.

\bibitem{romera2004}
\rd{Romera, R., Valdes, J.E., Zequeira, R.I., (2003).}
\newblock {Active-redundancy allocation in systems}.
\newblock {\em IEEE Transactions on Reliability,} \textbf{53},  313-318.

\bibitem{Samaniego1985}
\rd{Samaniego, F. J. (1985).} 
\newblock{On closure of the IFR class under formation
of coherent systems}. 
\newblock{\em{IEEE Transactions on Information
Theory,}} R-\textbf{34}, 69–72.



\bibitem{samaniego2007}
Samaniego, F. J. (2007).
\newblock {\textit{System signatures and their applications in engineering reliability}.}
\newblock {Springer Sciences+Business Media, LLC}, New York.


\bibitem{shayao1991}
\rd{Shanthikumar, J. G., Yao, D.D., 1991.
\newblock {Bivariate Characterization of some stochastic order relations}.
\newblock {\em Advances in Applied Probability,} \textbf{23},  642-659.}

\bibitem{shaked2007}
Shaked, M., Shanthikumar, J. G. (2007).
\newblock \textit{Stochastic Orders}.
\newblock { Springer-Verlag, New York}.

\bibitem{singmis1994}
\rd{Singh, H., Misra, N. (1994).
\newblock {On redundancy allocations in systems,} 
\newblock {\em Journal of Applied Probability}, \textbf{31}, 1004-1014.}

\bibitem{singsin1997}
\rd{Singh, H., Singh, R. S. (1997).
\newblock {Note: optimal allocation of resources to nodes of series systems with respect to failure rate ordering,} 
\newblock {\em Naval Research Logistics,} \textbf{44}, 147-152.}

\bibitem{valdes2003}
Valdes, J. E., Zequeira, R. I. (2003).
\newblock {On the optimal allocation of an active redundancy in a two-component series system,} 
\newblock {\em Statistics and Probability Letters,} \textbf{63}, 325-332.
\bibitem{youli2014} \rd{You, Y., Li, X. (2014). \newblock {On allocating redundancies to $k$-out-of-$n$ reliability systems}, \newblock {\em Applied Stochastic Models in Business and Industry,} \textbf{30}, 361-371.}

\end{thebibliography}
\end{document}